# Charge dynamics in near-surface, variable-density ensembles of nitrogen-vacancy centers in diamond

**Siddharth Dhomkar[1,\*], Harishankar Jayakumar[1,\*], Pablo R. Zangara[1], and Carlos A. Meriles[1,2,†]**

[1]*Dept. of Physics, CUNY-City College of New York, New York, NY 10031, USA.*

[2]*CUNY-Graduate Center, New York, NY 10016, USA.*

**ABSTRACT**: Although the spin properties of superficial shallow nitrogen-vacancy (NV) centers have been the subject of extensive scrutiny, considerably less attention has been devoted to studying the dynamics of NV charge conversion near the diamond surface. Using multi-color confocal microscopy, here we show that near-surface point defects arising from high-density ion implantation dramatically increase the ionization and recombination rates of shallow NVs compared to those in bulk diamond. Further, we find that these rates grow linearly — not quadratically — with laser intensity, indicative of single-photon processes enabled by NV state mixing with other defect states. Accompanying these findings we observe NV ionization and recombination in the dark, likely the result of charge transfer to neighboring traps. In spite of the altered charge dynamics, we show one can imprint rewritable, long-lasting patterns of charged-initialized, near-surface NVs over large areas, an ability that could be exploited for electrochemical biosensing or to optically store digital data sets with sub-diffraction resolution.

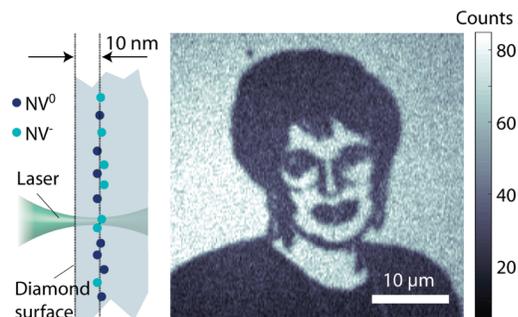

As one of the few paramagnetic point defects susceptible to optical spin initialization and readout under ambient conditions, the negatively charged nitrogen-vacancy (NV⁻) center is presently the subject of intense studies for applications in quantum information processing[1,2], various forms of micro- and nano-scale sensing[3-7], or as a platform for hyperpolarized magnetic resonance[8-10], among other examples. Since near-surface NVs are key to many of these applications, much effort has been devoted to identifying and understanding the pathways of surface-induced spin relaxation, and to engineering diamond surfaces where the impact of these mechanisms on the NV⁻ spin coherence times is reduced to a minimum.

Comparatively less attention, however, has been paid to the influence of surface proximity on the dynamics of NV charge state inter-conversion. Indeed, the charge instability of nano-diamond-hosted NVs — e.g., in the form of NV fluorescence blinking — has already been documented considerably[11], and its sensitivity to the environment has been exploited, for example, to probe electrochemical potentials in solution[12]. Unfortunately, the systematic characterization of surface effects on the charge dynamics of nano-crystalline NVs is challenging because the multi-faceted, often-irregular structure of diamond nanoparticles makes it difficult to ensure uniform surface termination. Further, other (often-prevailing) criteria of nano-diamond surface termination — including, e.g., reduced nano-particle aggregation or target-tailored surface functionalization — are difficult to reconcile with NV charge stability.

Recent ab-initio calculations of the energy level structure in bulk single-crystalline diamond expose the impact of surface proximity in all its complexity[13,14]. Specifically, it is found that typical diamond surfaces contain image and acceptor states with sub-bandgap energies that compromise the photo-stability of the NV through hybridization with the point defect gap states. The type and density of surface states strongly depends on the surface termination and hence can be controlled with the aid of suitable surface chemistry. Importantly, mixing with surface states adds to (but differs from) the so-called "band-bending", known to depend on the surface characteristics[15]. For example, hydrogen-terminated diamond features strongly negative electron affinity and upward band-bending, ultimately leading to *p*-type surface conductivity in the form of a two-dimensional (2D) hole gas. Therefore, hydrogenated surfaces deplete NV⁻ from its excess electron and make NV⁰ the dominant charge state. Oxidation partly removes band-bending, in which case both the NV⁻ and NV⁰ charge states are found to co-exist; positive electron affinity — i.e., downward band-bending — can be attained via surface fluorination[16], which has been shown to increase the fractional concentration of shallow NV⁻. Note, however, that the band-bending model alone is insufficient to explain processes such as blinking — associated with the appearance of acceptor surface states — and hence cannot be universally invoked to explain $NV^-$ discharge. We will return to this point later.

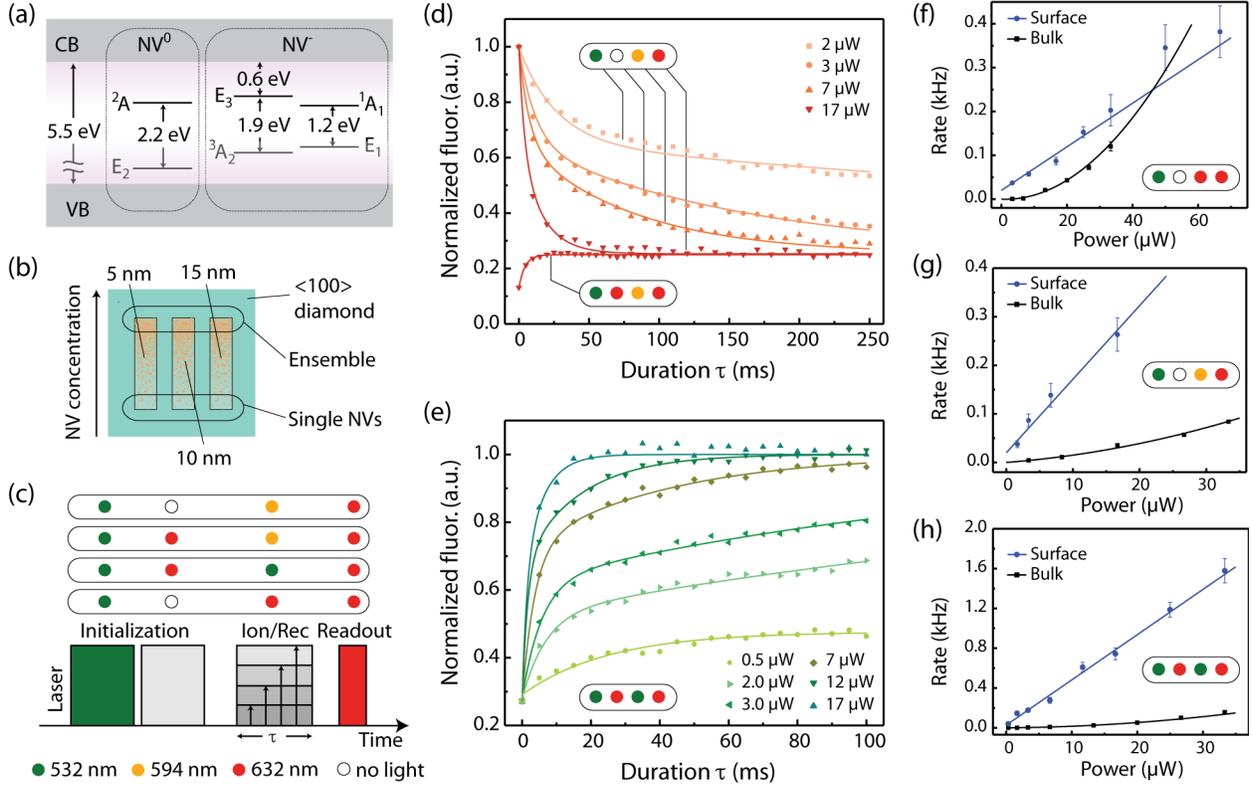

**Figure 1.** Comparing charge dynamics of shallow and bulk NVs. (a) Energy level diagram of the neutral and negatively-charged NVs. CB and VB denote the conduction and valence bands, respectively. The faded purple background near the bottom of the conduction band represents an additional set of localized and delocalized states present near the surface. (b) Schematics of the implanted diamond crystal featuring NVs at three different depths, 5, 10 and 15 nm. A concentration gradient allows us to address NV ensembles of variably size, down to individual color centers. (c) Alternative experimental protocols comprising NV charge initialization, excitation, and readout. The color labels above each laser pulse denotes the laser wavelength; empty circles indicate no light (and no wait time, i.e., the next light pulse follows immediately after). (d) Fluorescence from 10-nm-deep NVs upon 594 nm excitation of variable duration and intensity. The initial $NV^-$ population is maximum for the upper four traces and negligible for the lower trace. Solid traces indicate fits to double exponential curves; the color code denotes the laser pulse sequence as presented in (c). (e) Same as in (d) but for 532 nm excitation. (f) Ionization rates under 632 nm excitation for shallow (10 nm deep) and bulk $NV^-$. (g) Same as above but for 594 nm excitation. (h) 'Recharge' rates under green excitation for 10-nm-deep and bulk NVs. In (d) through (h) the power and duration of the 532 nm (632 nm) initialization pulse are 17 μW and 15 ms (17 μW and 100 or 300 ms), respectively; the duration and power of the 632 nm readout pulse are 17 μW and 500 μs, respectively.

Here we experimentally investigate the charge dynamics of near-surface NVs via multi-color confocal microscopy. Using bulk crystal NVs as a reference, we observe up to ten-fold faster charge conversion rates, depending on the illumination wavelength. We also find that the photo-ionization and recombination rates grow linearly, not quadratically, within the range of intensities probed (≲100 μW/μm²), indicative of distinct one-photon processes. Further, we show that both the negative and neutral charge states are unstable and slowly interconvert in the dark to attain an equilibrium concentration that depends on the implantation conditions. Long-term charge control of surface NVs is, nonetheless, possible as we show through the writing of arbitrary charge patterns seen to persist over several days.

The known charge dynamics of bulk NVs exposed to optical excitation can be understood with the aid of the energy diagrams in Fig. 1a: In the negatively charged state NVs ionize into $NV^0$ by ejecting an electron into the diamond conduction band via the absorption of one photon with energy >2.6 eV (i.e., 477 nm) or by the successive absorption of two photons with energies >1.946 eV (i.e., 637 nm).[17,18] In the latter mechanism, the first photon propels $NV^-$ to its first excited state and the second photon ejects the electron into the conduction band. Similarly, $NV^0$ can be photo-converted back to $NV^-$ either via a single photon absorption (for energies >2.94 eV (422 nm)) or by another two-step process involving the absorption of a photon with energy >2.156 eV (i.e., 575 nm) followed by the excitation of an electron from the diamond valence band. Relevant to the studies herein, these two-step, one-photon processes manifest experimentally in the form of a quadratic dependence on light intensity in both the ionization and recombination rates[18] (see below). Note that with the above constraints, red (632 nm) illumination results (mostly) in a one-directional charge conversion process, from NV$^-$ into NV$^0$, while the reverse process is largely suppressed. By contrast, green (532 nm) excitation dynamically modulates the NV charge state between negative and neutral; at this wavelength the equilibrium NV$^-$ population is approximately 75%.[18]

A schematic of the diamond sample we study herein is shown in Fig. 1b: Starting with an electronic grade, [100] diamond crystal (E6), we use focused nitrogen ion implantation followed by thermal annealing in vacuum to create shallow NV sets at three different average depths, namely, 5, 10, and 15 nm (see Methods); during implantation we gradually vary the ion fluence to create a nitrogen concentration gradient, from an (estimated) surface density of 4.2 ppm at its maximum, down to about 0.4 ppm (vertical strips in Fig. 1b). We estimate the NV concentration from the NV formation efficiency, ~1% for the present implantation conditions[19]. To favor the formation of negatively charged NVs, we partially oxygenate the diamond surface via acid boiling for hour-long periods.

Firstly, we compare the photo-ionization and recombination rates of near-surface and bulk NVs. In order to determine these rates we implement the set of protocols described in Fig. 1c: In all cases we use green excitation (15 ms, 17 μW) to bring the initial NV$^-$ population to a maximum, which we then optionally transform into mostly NV$^0$ via a red laser pulse (100 or 300 ms, 17 μW), depending on the desired initial charge state. This strategy circumvents ambiguities in the starting charge distribution arising from green- and red-induced nitrogen ionization[20]. NV charge readout is carried out with a red laser pulse (500 μs, 17 μW), whose amplitude and duration is chosen to make NV$^-$ ionization negligible. This form of readout allows us to derive in each case the fractional NV$^-$ population upon comparison of the observed fluorescence amplitude with that from a reference state (see Methods).

Focusing for now on the charge dynamics of 10-nm-deep, high-density ensembles, Figs. 1d and 1e respectively show the measured NV$^-$ fluorescence response after orange and green illumination of variable intensity and duration. In both cases we observe a bi-exponential evolution featuring a faster charge conversion at early times followed by a slower, asymptotic transformation towards an equilibrium charge concentration. The origin of this longer-term response with quadratic power dependence (Fig. S2 in the supplementary material) — also present in bulk samples, see Section I.C in the Supplementary Material — is presently unknown but we suspect it is connected to the saturation (or depletion) of available charge traps within the probed volume[21]. To avoid ambiguities, we characterize the charge dynamics in each case through the faster response at short times, governed by the smaller time constant in the bi-exponential fit.

Figs. 1f through 1h display the results corresponding to charge inter-conversion during red, orange, and green illumination, respectively; for comparison, we include the rates derived from similar observations in a reference Type 1b diamond (see Methods). Depending on the initial charge state — predominantly NV$^-$ or NV$^0$ in Figs. 1d and 1e, respectively — we observe a growth or a reduction of the NV$^-$ fluorescence with pulse duration; below we refer to the inverse of the (shorter) time constants in either case — effectively the initial slopes in the fluorescence traces — as the NV "ionization" and "recombination" rates, respectively. The latter is, of course, a simplification since both processes take place simultaneously[18], both for green and orange illumination (where significant NV$^0 \rightarrow$ NV$^-$ conversion is found to take place, see lower trace in Fig. 1d).

Compared to the response within the bulk crystal, shallow NVs tend to exhibit a faster charge photo-dynamics at low intensities (e.g., up to ten-fold in Fig. 1g). Further, we find that all NV charge conversion rates grow linearly with laser power, very much in contrast with bulk NVs where the dependence is quadratic. These observations hint at a fundamental change in the mechanisms governing the NV charge dynamics in the form of alternate ionization/recombination channels not present in bulk crystals. The linear dependence on light intensity indicates that one-photon processes — where the NV charge state changes upon absorption of a single photon — become dominant. The latter, in turn, is likely the result of mixing between the states of the NV and other localized (or delocalized) states originating from neighboring defects. For example, single-photon excitation of the NV electron into the hybrid excited states can result in the delocalization of the electron from the NV center.

The exact nature of these states in our sample is not fully clear. Recent ab-initio work[13] shows that typical diamond surfaces posses image and acceptor states with sub-bandgap energies that can significantly impact the charge dynamics of shallow color centers such as the NV. As the NV excited triplet hybridize (for example, through mixing with delocalized states), electron scattering away from the defect site upon photo-excitation can lead to single-photon NV$^-$ ionization[13]. In our case, however, surface states, though influential, do not seem to play a dominant role. A first indication is shown in Fig. 2a where we plot the NV$^-$ ionization rates as a function of laser power for different implantation depths. The observed dependence remains linear in all cases although the absolute rates tend to be smaller for deeper NVs. As the NV density decreases, however, the response becomes quadratic, even for the NVs closest to the surface (Fig. 2b). Further, for a given implantation depth, we find that the absolute rates tend to decrease with NV density (Fig. 2c), hence suggesting that state hybridization takes place through mixing with states produced during ion bombardment, for example, those from vacancy complexes forming during typical implantation and annealing protocols[22].

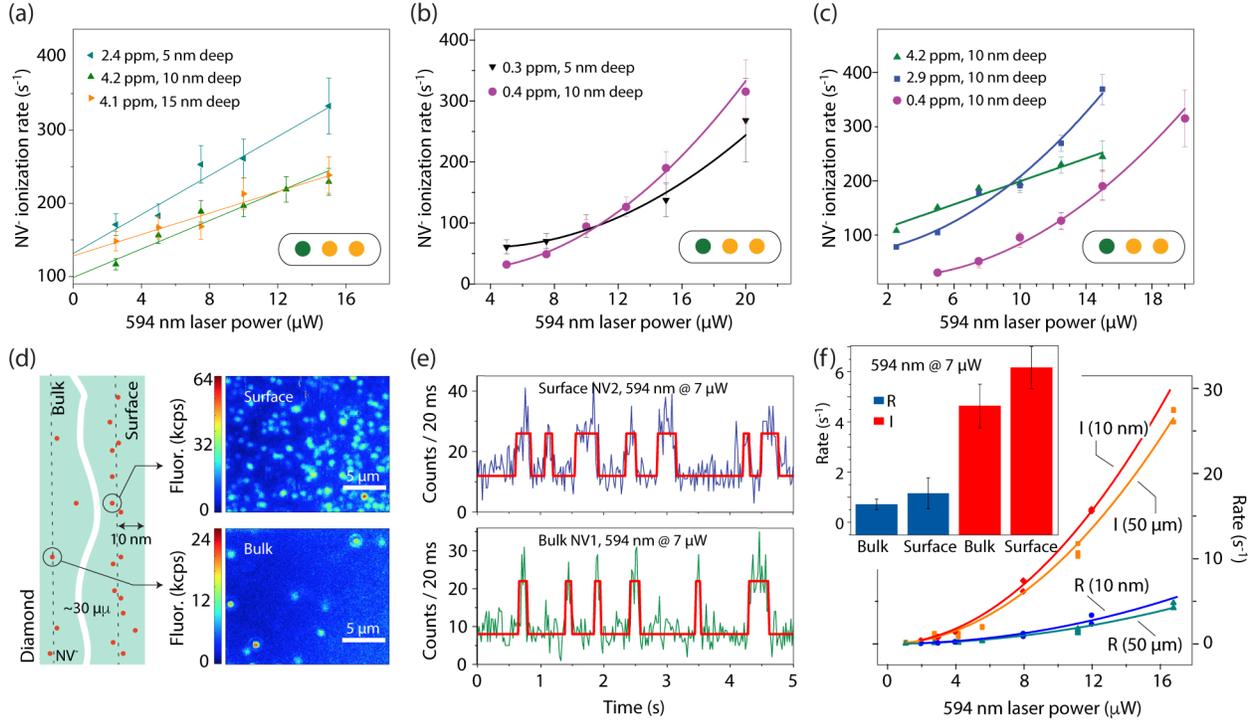

**Figure 2.** (a) NV⁻ ionization rates at three different depths for high-density NV ensembles. In all cases, we observe a linear response with laser power (solid lines are linear fits). (b) Same as in (a) but for the low density end of the implanted strips; ionization rates now grow quadratically (solid lines) with illumination intensity. (c) NV⁻ ionization rates at three different locations of variable NV⁻ content within the 10-nm-deep implanted strip. (d) Confocal fluorescence microscopy in the low density end of the 10-nm-deep strip allows us to identify individual NV⁻ emitters (top image). Diplacing the confocal plane to deeper within the bulk crystal, we can probe intrinsic (i.e., not implanted) NVs (lower image), which can then be used as a reference. (e) Fluorescence time trace from NVs on the surface (top) and bulk of the crystal (bottom) under cw 594 nm excitation (5 μW); blinking arises from dynamic NV charge ionization and recombination. (f) Comparative NV charge recombination (R) and ionization (I) rates for sample surface and bulk NVs (10 nm and 50 μm, respectively) as derived from (e). The insert shows statistics for a total of 20 NVs at 7 μW. In (a) through (c), ionization rates are determined following the protocol in Fig. 1g, except that the we use a 594 nm laser for readout (10 μW and 500 μs, lower right corner); the legend indicates the calculated NV concentration in parts per million (ppm) and depth.

Figs. 2d through 2f further support this idea: Here we compare the response from *individual* NVs either within the bulk of the crystal or near the low-density end of the 10-nm-deep strip (see confocal images in the lower and upper half of Fig. 2d, respectively). To determine the ionization and recombination rates, we record the fluorescence time trace during continuous 594 nm excitation, where conversion between the negative and neutral charge states shows in the form of blinking (Fig. 2e). Averaging over a set of individual bulk or shallow NVs (totaling 10 and 12, respectively), we find similar ionization and recombination rates for both groups (Fig. 2f), consistent with recent observations[23]. We surmise, therefore, that crystal imperfections rather than surface proximity are responsible for the observed phenomenology.

Unlike bulk NVs — where a given charge state remains stable over at least a week[24] (and possibly longer due to ill-defined Fermi levels[25]) — we find in this sample a gradual charge transformation in the dark, preferentially to the neutral state. Fig. 3 contains a summary of our observations: Following charge initialization via a 50 μW, 532 nm laser pulse, we probe the system evolution by comparing the NV⁻ fluorescence determined from 594 nm readout pulses immediately after the green pulse and following a dark time interval T (Fig. 3a). The upper half of Fig. 3b contains the results for various NV densities along the 10 nm deep implantation strip. In all cases the NV⁻ population progressively decays with time at a gradually diminishing rate, similar to a prior report[26]. We find a clear correlation between the local nitrogen implantation density and the fraction of negatively charged NVs transforming to neutral at a given time. Since the former arguably correlates with the local number of defects in the diamond lattice, our observations indicate that charge capture by trap states — a process distinctly different from band-bending — is responsible for the gradual discharge of NV⁻ in the dark.

We qualitatively reproduce our results via a kinetic Monte Carlo where the excess electron of a negatively charged NV at the center of a virtual diamond lattice randomly transfers to a neighboring trap (Fig. 3c). In our model we assume that the unit time probability characterizing the electron transfer decreases exponentially with the distance between the NV and the trap. The upper half in Fig. 3d shows the averaged results of our simulations for different trap densities, which we assume proportional to the local nitrogen concentration[27] (see Section II in the Supplementary Material). We attain reasonable agreement with the experimental observations (Fig. 3b), namely the calculated time dependence has the same overall shape featuring a fast decay followed by an increasingly slow evolution; as expected, the fraction of NV⁻ transforming into NV⁰ grows with the trap density.

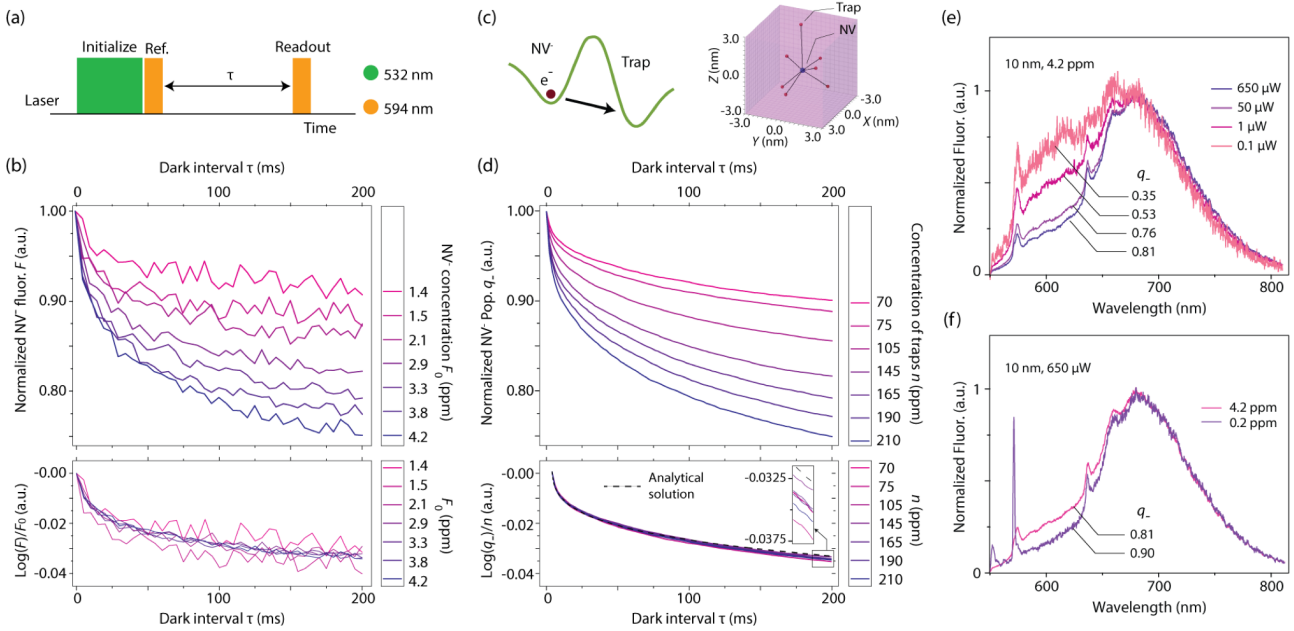

**Figure 3.** (a) Schematics of the experimental protocol. After initialization into NV⁻ via 532 nm laser pulse (50 μW, 10 ms), we determine the fractional NV⁻ population by comparing the fluorescence from a reference pulse to that of readout pulse following a dark interval of variable duration τ; the wavelength, power, and duration of both the reference and readout pulses are 594 nm, 10 μW and 500 μs, respectively. (b) (Top) NV⁻ fluorescence $F$ in the dark as a function of τ for various 10-nm-deep NV⁻ densities expressed in terms of $F_0$, the fluorescence at τ=0. (Bottom) Renormalized fluorescence, see text. (c) (Left) In the presence of neighboring traps, negatively charged NVs can transform into neutral via electron tunneling or thermal activation. (Right) To reproduce the observed dynamics, we use a kinetic Monte Carlo model comprising an NV center and a random distribution of charge traps over a (5 nm)³ virtual crystal lattice. (d) (Top) Averaged time response of the fractional NV⁻ population for a variable number of traps $N$ as calculated from kinetic Monte Carlo. (Bottom) Renormalized fractional NV⁻ population. Also shown for comparison is the analytical solution assuming the charge trap probability distribution is continuous over space. (e) Optical spectroscopy of 10-nm-deep NVs under 532 nm excitation for various laser intensities in the high-density region of the implantation strip. (f) Same as above but for fixed laser power (650 μW) and two different NV concentrations.

To gain a better understanding of the processes at play we write the NV⁻ fraction $q_-(t)$ at a given time $t$ as

$$q_-(t) \propto \langle \exp\left(-\sum_i t/\tau_{ij}\right)\rangle_j, \qquad (1)$$

where $1/\tau_{ij} \propto \exp(-kr_{ij})$ is the unit time probability of a charge transfer from the $j$-th NV⁻ to the $i$-th trap at a distance $r_{ij}$, $k$ is a constant, and brackets indicate average over all negatively-charged NVs. Assuming that traps are uniformly distributed with density $n$ and that NVs are sufficiently sparse (so that a given trap cannot be accessed by electrons from different NVs), Eq. (1) takes the form[28]

$$q_-(t) \propto \exp\left(n/n_{eff}\, g(t/\tau_{eff})\right), \qquad (2)$$

where $g(t/\tau_{eff})$ is a function of time, and $n_{eff}$ and $1/\tau_{eff}$ are defect-specific constants. Provided we take logarithm and divide the result in each case by the corresponding trap density $n$, it follows from Eq. (2) that all traces must collapse into a single 'universal' response. As shown in the lower half of Figs. 3b and 3d, our experimental and numerical results are consistent with this prediction. We also find reasonable agreement with an analytical formula[28] derived from assuming a continuous (i.e., non-discrete) distribution of traps (dashed black trace in the lower half of Fig. 3d), though the predicted NV⁻ population takes slightly higher values at longer times (see figure inset).

Interestingly, the observed charge transformation of the NVs was also captured through optical spectroscopy-based experiments. Here, the optical spectrum of the near-surface NVs was recorded with a continuous 532 nm illumination to determine the NV⁻ fraction. Since 532 nm illumination induces NV charge inter-conversion between negative and neutral, the fractional NV⁻ content under continuous optical excitation emerges from the interplay between the ionization and recombination rates as well as the NV⁻ charge transfer rate to neighboring traps. This is demonstrated in Fig. 3e where we collect the optical spectra from the high-density end of the NV strips for 532 nm illumination of variable intensity. For all implantation depths we find that the NV⁻ fraction invariably grows with laser power, starting from a minimum of ∼35% for 0.1 μW to ∼81% for ∼600 μW. The exact NV⁻ content is difficult to extract with accuracy because the NV⁻ contribution to the spectrum has a different shape relative to that observed in individual bulk NVs. In particular, the Stokes-shifted section of the spectrum has a greater relative size, which we interpret as an indication of additional phonon-mediated optical relaxation channels (see Section I.F in the Supplementary Material). Similar to the trend in Fig. 2, the NV⁻ fraction converges to bulk crystal values as the concentration of shallow NVs becomes lower (Fig. 3f).

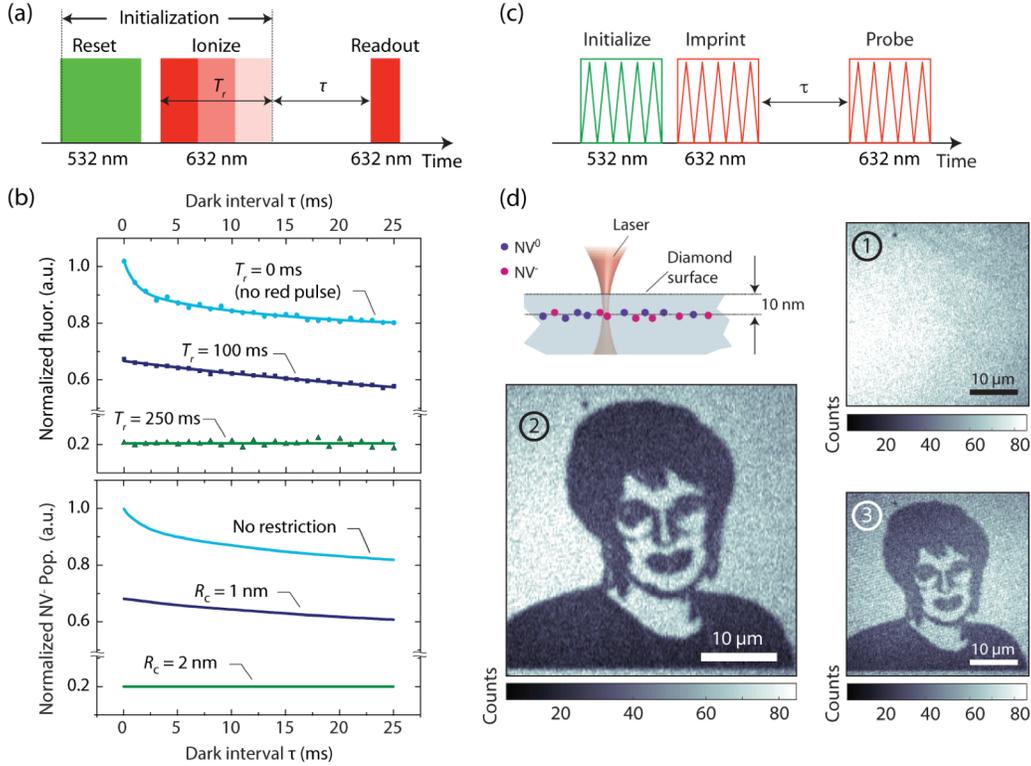

**Figure 4:** (a) Experimental protocol. Following 532 nm initialization (17 μW, 15 ms) and partial NV⁻ ionization via 632 nm illumination of variable duration (17 μW), we probe the fractional NV⁻ population via a 632 nm readout pulse (17 μW, 500 μs) after a dark time interval $T$. (b) (Top) Measured NV⁻ fluorescence as a function of $T$ for red ionization of different intensities, as shown in the figure labels; solid lines are guides to the eye. (Bottom) Calculated response as determined from a kinetic Montecarlo simulation assuming a trap density of 210 ppm. The upper, light blue trace reproduces the calculation in Fig. 2d for the corresponding time window; the lower traces use the exact same model except that we assume no traps are available within a sphere of critical radius $R_c$ (1 and 2 nm for the middle and lower traces, respectively). In each case, the initial charge state of the NV is chosen to match the measured NV⁻ fractional population at zero delay. (c) Charge imprinting protocol; zig-zags indicate laser scan. (d) Proof-of-principle data encoding in the high-density (420 ppm N concentration) segment of the on 10-nm-deep implantation strip. Following initialization via five 532 nm scans (image 1 in the upper right) we selectively ionize shallow NV⁻ to create a pre-defined charge pattern (lower left). Images 2 and 3 show the result via a 632 nm readout scan after a wait time $T$ of one minute and 15 hours, respectively. The 532 nm laser power and illumination time per pixel is 17 μW and 2 ms, respectively; the power and illumination time during 632 nm NV⁻ ionization is 17 μW and 0-20 ms, respectively. In all cases, readout is carried out via a 632 nm scan (17 μW, 2 ms integration time per pixel).

Interestingly, we find that the NV⁻ charge dynamics in the dark depends on the illumination history. Fig. 4a introduces an example protocol where 532 nm initialization is followed by partial 632 nm bleach. Compared to the case where no red pulse is applied (upper light blue trace), we find a modified, slower charge conversion (middle dark blue trace), which we rationalize in terms of an altered microscopic charge environment. More specifically, since the faster changing segment in the $q_-(t)$ curve can be related to the shorter charge transfer times (see Eq. (1)), we can qualitatively reproduce our observations by imposing a critical radius $R_c$ below which no traps are available (see simulated traces in the lower half of Fig. 4b). Physically, the latter amounts to assuming that red illumination preferentially transfers the excess NV⁻ electron to the nearest empty trap — perhaps via photo-induced tunneling from the excited state. We warn the above model must be taken with caution, particularly given the incomplete knowledge of the potentials driving the charge transfer process (see below). For future reference, we note that the NV⁻ fluorescence slightly tends to grow after a near-full bleach (green trace in the upper half of Fig. 4b), thus hinting at a non-negligible NV⁻ equilibrium population in the dark.

The physical nature of the charge traps at play is presently unclear but we hypothesize vacancy complexes play a dominant role. Among the latter, the di-vacancy V₂ — a defect forming efficiently via diffusion of single vacancies during typical sample annealing protocols[22] — lists as the most likely candidate. Ab initio calculations[22] indicate the ground state of both the neutral and negatively charged V₂ lie below that of NV⁻, hence making electron transfer to these defects energetically favorable. Another relevant defect is the N$_s$–V–N$_s$ complex, also forming during annealing in nitrogen-rich diamond and potentially able to adopt a negatively charged state via electron capture from NV⁻. For completeness, we note that while neutral vacancies could also serve as electron traps, its characteristic zero-phonon-line at 741 nm is absent in the optical spectra from this sample (see below) and can, therefore, be ruled out.

Finally, we turn our attention to the long-term charge conversion dynamics of NVs, a subject relevant to applications in electric field sensing[29] (where multi-second-long processes are not uncommon), and for data storage[24] (where charge stability over several days or more is a requisite). The latter use is particularly intriguing because sub-diffraction charge control — e.g., with the aid of near-field scanning microscopy[30] — could be exploited to create ultra-dense optical memories. As a first step in this direction, we implement the protocol of Fig. 4c: Upon initializing NVs in the negatively charged state over a large area (~40×40 μm$^2$ in the high-density end of the 10-nm-deep implantation strip), we selectively photo-ionize NV⁻ into NV⁰ via 632 nm light so as to encode an arbitrary charge pattern (in this case, the well-known portrait of Richard Feynman). We probe the result via a quick 632 nm laser scan following a variable dark time interval $T$. Fig. 4d displays images of the area of interest after charge initialization and encoding. Using as a reference the result obtained minutes after encoding (image 2 in Fig. 4d), we observe, as expected, preferential transformation into NV⁰ within the NV⁻-rich areas; dark regions of the pattern, on the other hand, exhibit non-negligible back-conversion into NV⁻. Despite these ongoing processes, we attain good contrast even for dark intervals exceeding 15 hours (image 3 in Fig. 4d).

In summary, our experiments show that shallow NVs exhibit singular charge dynamics in the form of single-photon ionization and recombination, and charge inter-conversion in the dark. The impact of these processes on the system response correlates with the nitrogen content, stronger in areas of higher ion implantation density. Since the charge dynamics of isolated shallow and bulk NVs is similar, we conclude the phenomena observed in ensembles arise from the presence of other defects created during ion implantation and sample annealing, possibly substitutional nitrogen and di-vacancies. In the limit of high defect concentrations, one can rationalize single-photon NV⁻ ionization as a manifestation of inter-defect proximity effects, e.g., in the form of hybridization of the NV⁻ excited states with trap states. Depending on the target application, state mixing with neighboring traps could be exploited, for instance, to enhance the selectivity of spin-to-charge conversion protocols, relevant to nanoscale sensing.

Future work will therefore be necessary to unambiguously identify the physical nature of these defects and better control their effect on neighboring NVs. For example, optimized NV engineering protocols — including delta-doping[31,32], and annealing at higher temperatures[22,33] or upon overgrowth of a boron-doped layer[34] — may help stabilize the charge state of shallow NVs over longer time windows, of interest for data storage. Alternatively, surface termination could be tailored to make NV charge conversion sensitive to electrochemical processes on the diamond surface. By the same token, experiments as a function of temperature will be useful to separately assess the contribution of tunneling and thermal activation to the electron transfer affecting NV⁻ in the dark.

## ASSOCIATED CONTENT

**Supporting Information:** Contains additional information on sample characteristics and instrumentation used. It also provides additional details on the Monte Carlo modelling.

## AUTHOR INFORMATION

**Corresponding author:**
†E-mail: cmeriles@ccny.cuny.edu
**Notes**
(*) denotes equally contributing authors.
The authors declare no competing financial interests.

## ACKNOWLEDGMENTS

We thank Marcus Doherty, Audrius Alkauskas, for fruitful discussion, and Jacob Henshaw for assistance with some of the experiments. We also thank Prof. Vinod Menon for providing access to a high-resolution optical spectrometer. All authors acknowledge support from the National Science Foundation through grants NSF-1619896 and NSF-1401632, and from Research Corporation for Science Advancement through a FRED Award. The authors acknowledge the facilities and research infrastructure support of the NSF CREST-IDEALS, NSF grant number HRD-1547830.